# Deterministic Construction of Binary, Bipolar and Ternary Compressed Sensing Matrices

Arash Amini, and Farokh Marvasti, *Senior Member, IEEE*

*Abstract*—In this paper we establish the connection between the Orthogonal Optical Codes (OOC) and binary compressed sensing matrices. We also introduce deterministic bipolar $m \times n$ RIP fulfilling $\pm 1$ matrices of order $k$ such that $m \leq \mathcal{O}\big(k(\log_2 n)^{\frac{\log_2 k}{\ln \log_2 k}}\big)$. The columns of these matrices are binary BCH code vectors where the zeros are replaced by $-1$. Since the RIP is established by means of coherence, the simple greedy algorithms such as Matching Pursuit are able to recover the sparse solution from the noiseless samples. Due to the cyclic property of the BCH codes, we show that the FFT algorithm can be employed in the reconstruction methods to considerably reduce the computational complexity. In addition, we combine the binary and bipolar matrices to form ternary sensing matrices ($\{0, 1, -1\}$ elements) that satisfy the RIP condition.

## I. INTRODUCTION

MINIMIZATION of the number of required samples for unique representation of *sparse* signals has been the subject of extensive research in the past few years. The field of compressed sensing, which is originated by the pioneering works in [1], [2], [3] deals with the reliable reconstruction of an $n \times 1$ but $k$-sparse vector $\mathbf{x}_{n \times 1}$ from its linear projections ($\mathbf{y}_{m \times 1}$) onto an $m$-dimensional ($m \ll n$) space: $\mathbf{y}_{m \times 1} = \mathbf{\Phi}_{m \times n} \mathbf{x}_{n \times 1}$. The two main concerns in compressed sensing are 1) selecting the sampling matrix $\mathbf{\Phi}_{m \times n}$ and 2) reconstructing $\mathbf{x}_{n \times 1}$ from the measurements $\mathbf{y}_{m \times 1}$ by exploiting the sparsity constraint.

The sampling matrix is usually treated by random selection of the entries; among the well-known random matrices are i.i.d Gaussian [1] and Rademacher [4] matrices. In general, the exact solution to the second concern, is shown to be an NP-complete problem [5]; however, if the number of samples ($m$) exceeds the lower bound of $m > \mathcal{O}(k \log(n/k))$, $\ell_1$ minimization (Basis Pursuit) can be performed instead of the exact $\ell_0$ minimization (sparsity constraint) with the same solution for almost all the possible inputs [2], [5]. There are also greedy techniques such as Matching Pursuit method [6], [7] that can be used.

In this paper we are interested in deterministic as opposed to random sampling (sensing) matrices. Deterministic sampling matrices are useful because in practice, the sampler has to be a deterministic matrix; although random matrices perform quite well on the average, there is no guarantee that a specific realization works. Moreover, by proper choice of the matrix, we might be able to improve some features such as computational complexity and compression ratio.

In deterministic designs, one of the well-studied conditions on the sensing matrix which guarantees stable recovery for a number of reconstruction methods, is the so called Restricted Isometry Property (RIP) [2]: we say that the matrix $\mathbf{A}_{m \times n}$ obeys RIP of order $k$ with constant $0 \leq \delta_k < 1$ (RIC) if for all $k$-sparse vectors $\mathbf{x}_{n \times 1}$, we have:

$$1 - \delta_k \leq \frac{\|\mathbf{A}\mathbf{x}\|_{\ell_2}^2}{\|\mathbf{x}\|_{\ell_2}^2} \leq 1 + \delta_k$$

The basis pursuit and greedy methods can be applied for recovery of $k$-sparse vectors from noisy samples with good results if the matrix $\mathbf{A}$ obeys RIP of order $2k$ with a good enough constant $\delta_{2k}$ [7], [8].

For the deterministic approaches, the Vandermond matrices might seem to be good options at the first glance; any $k$ columns of a $k \times n$ Vandermond matrix are linearly independent. Thus, after normalizing the columns, the matrix satisfies the RIP condition of order $k$ (only the left inequality). In other words, arbitrary RIP-constrained matrices could be constructed in this way; however, when $n$ increases, the constant $\delta_k$ rapidly approaches 1 and some of the $k \times k$ submatrices become ill-conditioned [9] which makes these matrices impractical. Among the proposed and relatively successful deterministic schemes are complex-valued $m \times m^2$ chirp-based matrices [10]; although, they are not supported with any established RIP order, it is shown that the combinatorial $\ell_0$ minimization problem perfectly recovers the original sparse vector (sparsity order below a threshold) from noiseless samples. A connection between the coding theory and sensing matrices is established in [11] where second order Reed-Muller codes are used to construct bipolar ($\pm 1$) $2^l \times 2^{\frac{l(l+1)}{2}}$ matrices but similar to the chirp-based case, they lack a guarantee on the RIP order. The very simple matrices for which an RIP order can be established are those formed by concatenating two incoherent unitary matrices such as the so called Spikes and Sines; this technique results in $m \times 2m$ matrices that satisfy RIP of order $\lfloor \sqrt{m} \rfloor + 1$. More general than the concatenation approach, the incoherence (small inner product between distinct columns) can be used to establish RIP for the matrices constructed by Grassmannian or equiangular tight frames [12]; unfortunately, although there are almost sharp conditions for the existence of such matrices [13], their explicit construction is only known for $m \times n$ matrices with $\frac{n}{m} \leq 2$ [14]. Furthermore, no matter how small the required RIP order is, the parameter $n$ is upper-

A. Amini and F. Marvasti are with the Advanced Communication Research Institute (ACRI), department of Electrical Engineering, Sharif University of Technology, Tehran, Iran e-mail: arashsil@ee.sharif.edu , marvasti@sharif.edu.
Manuscript received Aug. 19, 2009; revised July 23, 2010.
This work is supported by the Iranian Telecommunication Research Center (ITRC).

bounded by $\binom{m+1}{2}$ in these matrices; i.e., the increasing rate of $n$ is at most quadratic with respect to $m$. Devore's binary $p^2 \times p^{r+1}$ matrices are among the very few deterministic designs which provide RIP without restricting the growth of $n$ to a quadratic function of $m$; here, $p$ should be a prime power and $kr < p$ where $k$ is the desired RIP order [15]. Another binary matrix construction with $m = k2^{\mathcal{O}(\log \log n)^E}$ measurements ($E > 1$) is investigated in [16] which employs hash functions and extractor graphs. Recently, almost bound-achieving matrices have been proposed in [17] which, rather than the exact RIP, satisfy the statistical RIP (RIP inequalities hold with high probability if the support of the $k$-sparse vector is drawn uniformly at random from all the $\binom{n}{k}$ possibilities).

Since the deterministic designs are mainly motivated by the capability of being implemented, practical aspects such as sensing and reconstructing procedures should be taken into account. For this reason, we focus on the matrices that are composed of $0, \pm 1$, i.e., the elements that facilitate the matrix multiplication (sensing). In addition, we design the matrices such that the simple greedy reconstruction methods (e.g., matching pursuit) can recover the sparse inputs from the compressed measurements. The main contributions of our paper are listed below:

1) We establish the connection between the optical codes and the binary sampling matrices. Using the results in the optical codes, we give a tight upper-bound on the number of columns in the binary sensing matrices and we show that the Devore's matrices are almost optimal.

2) Using the linear binary block codes (specifically, BCH codes) we introduce $(2^l - 1) \times 2^{\mathcal{O}\left(2^{(l-j)}\frac{\ln j}{j}\right)}$ bipolar ($\pm 1$) sensing matrices which obey the RIP of order $k \leq 2^j + 1$ ($l > j$); similar to Devore's design, the growth of $n$ in these matrices is not restricted to a quadratic function of $m$. Although these matrices have almost the same asymptotic sizes as the Devore's design (for similar RIP orders), for practical cases, we observe that the bipolar matrices satisfy higher RIP orders (by means of coherence) for the similar values of $m$ and $n$. In addition, since the bipolar design is based on the cyclic codes, the reconstruction algorithm can be expedited by exploiting the FFT algorithm.

3) By combining the binary and bipolar matrices, we generate ternary matrices with $m = p^2$ and $n = p^{r+1} 2^{\mathcal{O}\left(r \frac{\ln(\log_2 p - \log_2 r)}{\log_2 p - \log_2 r}\right)}$ which satisfy RIP of order $k < \frac{p}{r}$, where $p$ is a Mersenne prime. Note that the order of $n$ is slightly greater than that of the Devore's design and as we show, Devore's design is nearly optimal in binary schemes. Hence, to the best of our knowledge, the introduced ternary matrices have the largest order of $n$ for the same value of $m$ among the deterministic designs which guarantee the RIP of order $k$.

The rest of the paper is organized as follows: binary sampling matrices including the OOC codes and the discussion about the optimality of the Devore's design are discussed in the next section. In section III, the construction of bipolar matrices using block codes is studied. We combine the binary and bipolar matrices in section IV to form ternary matrices and we present the recovery methods to obtain the sparse signal from the measurements in section V. Section VI represents the numerical simulations and finally, section VII concludes the paper.

## II. BINARY SAMPLING MATRICES

In this section, we first introduce the approach for producing RIP-fulfilling matrices. The approach is similar to [15] and is not restricted to the binary matrices.

Let $\mathbf{A}_{m \times n}$ be a real matrix with normalized columns such that the absolute value of the inner product of each two columns does not exceed $\lambda$. Let $\mathbf{B}_{m \times k}$ be any matrix composed of $k$ distinct columns of $\mathbf{A}$ and define the Grammian matrix $\mathbf{G}_{k \times k} = \mathbf{B}^T \mathbf{B}$. Since the columns of $\mathbf{B}$ are normal, the diagonal elements of $\mathbf{G}$ are all equal to $1$; moreover, the absolute values of the non-diagonal elements of $\mathbf{G}$ do not exceed $\lambda$. Therefore, we have:

$$\forall \, 1 \leq i \leq k : \sum_{j \, , \, j \neq i} |g_{i,j}| \leq \lambda(k-1) \quad (1)$$

From the Gershgorin circle theorem, we know that the eigenvalues of $\mathbf{G}$ lie in the interval $[1 - \lambda(k-1) \, , \, 1 + \lambda(k-1)]$; thus, if $k$ is small enough such that $\delta_k = \lambda(k-1) < 1$, the matrix $\mathbf{A}_{m \times n}$ satisfies RIP of order $k$ with the constant $\delta_k$. In other words, in order to construct a sampling matrix for compressed sensing, we introduce $m \times n$ matrices for which we have $\lambda < \frac{1}{k-1}$.

Binary sampling matrices are RIP-fulfilling matrices with $0, 1$ elements prior to column normalization. A subset of such matrices was previously studied in the field of Optical Code Division Multiple Access (OCDMA) with the name of OOC [18]; since in the optical communication only positive values can be transmitted, each user is assigned a binary vector (signature) with a fixed weight (number of $1$'s) where the inner product of different vectors are small compared to the weight (in contrast to what OOC stands for, the signatures are not orthogonal). A useful upper bound (not necessarily achievable) for the maximum number of such binary vectors is given in [19]: if $R(m, w, \lambda)$ stands for the maximum number of $m \times 1$ binary vectors with weight $w$ such that the inner product of each two does not exceed $\lambda$ ($\lambda \in \mathbb{Z}$), we have:

$$R(m, w, \lambda) \leq \left\lfloor \frac{m}{w} \left\lfloor \frac{m-1}{w-1} \left\lfloor \dots \left\lfloor \frac{m-\lambda}{w-\lambda} \right\rfloor \dots \right\rfloor \right\rfloor \right\rfloor \quad (2)$$

where $\lfloor x \rfloor$ represents the largest integer not greater than $x$. Although the small value of the inner product of the signatures is the main key for proper detection of the communicated message in a multi-access scenario, in asynchronous cases, the circular cross correlation (inner product of a signature with the circularly shifted versions of another) and autocorrelation (inner product of a signature with its circularly shifted versions) are as important. Therefore, instead of a simple $\lambda$, two parameters are involved: $\lambda_a$ denotes the maximum value of the circular auto-correlation among all the code vectors when at least one and at most $m-1$ units of shift are applied and $\lambda_c$ denotes the maximum value of the circular cross-correlation among all the pairs. The OOC vectors are characterized by $(m, w, \lambda_a, \lambda_c)$; nonetheless, it is possible that the two correlation parameters are equal ($\lambda_a = \lambda_c = \lambda$), in this case, the OOC is referred to as a $(m, w, \lambda)$-code.

Now let $\mathcal{A}$ be a set of OOC vectors of size $m$ with weight $w$ and the correlation parameters $\lambda_a = \lambda_c = \lambda$; we also include all the possible circularly shifted versions of the codes in $\mathcal{A}$. According to the definition of OOC's, the inner product of each pair in $\mathcal{A}$ is upper bounded by $\lambda$. We construct the matrix $\mathbf{A}_{m \times n}$ by the normalized versions of the column vectors in $\mathcal{A}$ where $n = |\mathcal{A}|$ (the order of the columns is unimportant). With respect to the upper bound on the inner product of the vectors in $\mathcal{A}$, it is easy to verify that the matrix $\mathbf{A}$ satisfies RIP of order $k < 1 + \frac{w}{\lambda}$. Below we will only discuss one of the OOC designs using Galois fields [20]:

Let $q = 16^a$ where $a \in \mathbb{N}$ and let $\mathbb{F} = GF(q)$ with the primitive root $\alpha$. It is clear that $5 | q - 1$ which confirms the existence of an integer $d$ such that $q = 5d + 1$. Define:

$$\mathcal{D}_i = \{\alpha^{d+i}, \alpha^{2d+i}, \ldots, \alpha^{5d+i}\} \quad, \quad 0 \leq i \leq d - 1 \qquad (3)$$

Since the number of 1's are usually far less than that of 0's, it is common to represent the OOC vectors by their nonzero locations. For the above design, the length of the codes is equal to $q - 1$ ($m = 16^a - 1$) and the nonzero locations of each code vector is given in $\{\mathcal{C}_i\}_{i=1}^{d-1}$:

$$\mathcal{C}_i = \log_\alpha(\mathcal{D}_i - 1) \quad, \quad 1 \leq i \leq d - 1 \qquad (4)$$

Note that $1 \in \mathcal{D}_0$ and $\mathcal{D}_0$ is not used for code construction. It is shown in [20] that the above method produces $\frac{16^a - 6}{5}$ OOC vectors with the characteristic $(16^a - 1, 5, 2)$. Since in the construction of the sampling matrix for compressed sensing we include the circular shifts of the OOC vectors, we obtain a matrix with the size $(16^a - 1) \times n$ where $n \lessapprox \frac{(16^a - 1)(16^a - 6)}{5}$ for the RIP order of $k = 1 + \lfloor \frac{5}{2} \rfloor = 3$ and the constant $\delta_3 = 1 - \frac{5 - 2*(k-1)}{5} = 0.8$. In [20], by employing the same approach, other $(n, w, 2)$ OOC vectors with larger $w$'s (larger $k$'s in our case) are introduced which are claimed to be optimal considering the Johnson's inequality (2).

A matrix design independent of OOC codes is given in [15] that constructs $p^2 \times p^{r+1}$ binary matrices with the column weight of $p$ (prior to normalization) such that the inner product of each two columns does not exceed $r$ ($\frac{r}{p}$ after normalization). Here $p$ is a power of a prime integer; the matrix construction is based on polynomials in $GF(p)$. Although the introduced matrices do not achieve the bound predicted by the theory of random compressed sensing, using (2), we show that these structures are asymptotically optimal when $\frac{p}{r^2} \to \infty$:

$$\begin{aligned}
\lim_{\frac{p}{r^2} \to \infty} \frac{p^{r+1}}{R(p^2, p, r)} &\geq \lim_{\frac{p}{r^2} \to \infty} \prod_{i=0}^{r} \frac{p(p-i)}{p^2 - i} \\
&\geq \lim_{\frac{p}{r^2} \to \infty} \left(\frac{p(p-r)}{p^2 - r}\right)^{r+1} \\
&\geq \lim_{\frac{p}{r^2} \to \infty} \left(1 - \frac{r}{p}\right)^{r+1} \\
&= \lim_{\frac{p}{r^2} \to \infty} \left(\left(1 - \frac{r}{p}\right)^{-\frac{p}{r}}\right)^{-\frac{r(r+1)}{p}} \\
&\geq \lim_{\frac{p}{r^2} \to \infty} e^{-\frac{r(r+1)}{p}} = e^0 = 1 \qquad (5)
\end{aligned}$$

Besides, using (2), it can be shown that binary matrices are in general unable to reach the predicted compressed sensing bound unless $w = \mathcal{O}(m)$.

### III. BIPOLAR MATRICES VIA LINEAR CODES

In this section, we will describe the connection between the sampling matrix and coding theory[1]. Since the parameters $k, n$ are used in both compressed sensing and coding field, we distinguish the two by using the $\tilde{}$ notation for coding parameters; e.g., $\tilde{n}$ refers to the code length while $n$ denotes the number of columns of the sampling matrix.

Let $\mathcal{C}(\tilde{n}, \tilde{k}; 2)$ be a linear binary block code and $\mathbf{1}_{\tilde{n} \times 1}$ be the all 1 vector. We say $\mathcal{C}$ is 'symmetric' if $\mathbf{1}_{\tilde{n} \times 1} \in \mathcal{C}$. For symmetric codes, if $\mathbf{a}_{n \times 1}$ is a code vector, due to the linearity of the code, the complement of $\mathbf{a}_{n \times 1}$ defined as $\mathbf{a}_{n \times 1} \oplus \mathbf{1}_{\tilde{n} \times 1}$, is also a valid code vector; therefore, code vectors consist of complement couples.

***Theorem 1:*** Let $\mathcal{C}(\tilde{n}, \tilde{k}; 2)$ be a symmetric code with the minimum distance $\tilde{d}_{min}$ and let $\mathbf{A}_{\tilde{n} \times 2^{\tilde{k}-1}}$ be the matrix composed of code vectors as its columns such that from each complement couple, exactly one is selected. Define:

$$\mathbf{A}_{\tilde{n} \times 2^{\tilde{k}-1}} \triangleq \frac{1}{\sqrt{\tilde{n}}} \left(2\tilde{\mathbf{A}}_{\tilde{n} \times 2^{\tilde{k}-1}} - (1)_{\tilde{n} \times 2^{\tilde{k}-1}}\right) \qquad (6)$$

Then, $\mathbf{A}$ satisfies the RIP with the constant $\delta_k = (k-1)\left(1 - 2\frac{\tilde{d}_{min}}{\tilde{n}}\right)$ for $k < \frac{\tilde{n}}{\tilde{n} - 2\tilde{d}_{min}} + 1$ ($k$ is the RIP order).

**Proof.** First note that the columns of $\mathbf{A}$ are normal. In fact $2\tilde{\mathbf{A}}_{\tilde{n} \times 2^{\tilde{k}-1}} - (1)_{\tilde{n} \times 2^{\tilde{k}-1}}$ is the same matrix as $\tilde{\mathbf{A}}$ where zeros are replaced by $-1$ (bipolar representation); hence, the absolute value of each element of $\tilde{\mathbf{A}}$ is equal to $\frac{1}{\sqrt{\tilde{n}}}$ which reveals that the columns are normal.

To prove the RIP, we use a similar approach to that of [15]; we show that for each two columns of $\mathbf{A}$, the absolute value of their inner product is less than $\frac{\tilde{n} - 2\tilde{d}_{min}}{\tilde{n}}$. Let $\mathbf{a}_{\tilde{n} \times 1}, \mathbf{b}_{\tilde{n} \times 1}$ be two distinct columns of $\mathbf{A}$ and $\tilde{\mathbf{a}}_{\tilde{n} \times 1}, \tilde{\mathbf{b}}_{\tilde{n} \times 1}$ be their corresponding columns in $\tilde{\mathbf{A}}$. If $\tilde{\mathbf{a}}$ and $\tilde{\mathbf{b}}$ differ at $l$ positions, we have:

$$\langle \mathbf{a}, \mathbf{b} \rangle = \frac{1}{\tilde{n}}\left(1 \times (\tilde{n} - l) + (-1) \times l\right) = \frac{\tilde{n} - 2l}{\tilde{n}} \qquad (7)$$

Moreover, $\tilde{\mathbf{b}}$ and $\tilde{\mathbf{a}} \oplus \mathbf{1}_{\tilde{n} \times 1}$ (complement of $\tilde{\mathbf{a}}$) differ at $\tilde{n} - l$ positions and since all the three vectors $\{\mathbf{a}, \tilde{\mathbf{a}} \oplus \mathbf{1}_{\tilde{n} \times 1}, \mathbf{b}\}$ are different code words (from each complement couple, exactly one is chosen and thus $\mathbf{b} \neq \tilde{\mathbf{a}} \oplus \mathbf{1}_{\tilde{n} \times 1}$), both $l$ and $\tilde{n} - l$ should be greater than or equal to $\tilde{d}_{min}$, i.e.,

$$\begin{cases} l \geq \tilde{d}_{min} \\ \tilde{n} - l \geq \tilde{d}_{min} \end{cases} \Rightarrow \tilde{d}_{min} \leq l \leq \tilde{n} - \tilde{d}_{min}$$

$$\Rightarrow |\tilde{n} - 2l| \leq \tilde{n} - 2\tilde{d}_{min} \qquad (8)$$

---

[1]At the time of submitting this paper, we realized that the same connection is recently established in "On Random Construction of a Bipolar Sensing Matrix with Compact Representation," *IEEE Inf. Theo. Workshop, 2009* by Tadashi Wadayama. However, our work has been carried out independently and concurrently and we focus on the deterministic design of the codes rather than the probabilistic structure.

Note that $\mathbf{0}_{\tilde{n}\times 1}, \mathbf{1}_{\tilde{n}\times 1} \in \mathcal{C}$ and for each code vector $\mathbf{a}$, either $d(\mathbf{0}_{\tilde{n}\times 1}, \mathbf{a})$ or $d(\mathbf{1}_{\tilde{n}\times 1}, \mathbf{a})$ cannot exceed $\frac{\tilde{n}}{2}$; therefore, $\tilde{n} - 2\tilde{d}_{min} \geq 0$. Combining (7) and (8), we have:

$$|\langle \mathbf{a}, \mathbf{b} \rangle| \leq \frac{\tilde{n} - 2\tilde{d}_{min}}{\tilde{n}} \quad (9)$$

which proves the claim on the inner product of the columns of $\mathbf{A}$. This result together with Gershgorin circle theorem discussed in the previous section, proves the theorem ∎

The above theorem is useful only when $\tilde{d}_{min}$ is close to $\frac{\tilde{n}}{2}$ (denominator for the upper bound of $k$), which is not the case for the common binary codes. In fact, in communication systems, parity bits are inserted to protect the main data payload, i.e., $\tilde{k}$ bits of data are followed by $\tilde{n} - \tilde{k}$ parity bits. In this case, we have $\tilde{d}_{min} \leq \tilde{n} - \tilde{k} + 1$; thus, to have $\tilde{d}_{min} \approx \frac{\tilde{n}}{2}$, the number of parity bits should have the same order as the data payload which is impractical. In the next section we show how these types of codes can be designed using the well-known BCH codes[2].

### A. BCH codes with large $\tilde{d}_{min}$

Since the focus in this section is on the design of BCH codes with large minimum distances, we first briefly review the BCH structure.

BCH codes are a class of cyclic binary codes with $\tilde{n} = 2^{\tilde{m}} - 1$ which are produced by a generating polynomial $g(x) \in GF(2)[x]$ such that $g(x)|x^{2^{\tilde{m}}-1} + 1$ [21]. According to a result in Galois theory, we know:

$$x^{2^{\tilde{m}}-1} + 1 = \prod_{\substack{r \in GF(2^{\tilde{m}}) \\ r \neq 0}} (x - r) \quad (10)$$

Hence, the BCH generating polynomial can be decomposed into the product of linear factors in $GF(2^{\tilde{m}})[x]$. Let $\alpha \in GF(2^{\tilde{m}})$ be a primitive root of the field and let $\alpha^i$ be one of the roots of $g(x)$. Since $g(x) \in GF(2)[x]$, all conjugate elements of $\alpha^i$ (with respect to $GF(2)$) are also roots of $g(x)$. Again using the results in Galois theory, we know that these conjugates are different elements of the set $\{\alpha^{i 2^j}\}_{j=0}^{m-1}$. In addition, since $\alpha^{2^{\tilde{m}}-1} = 1$, $i_1 \equiv i_2 \pmod{2^{\tilde{m}} - 1}$ implies $\alpha^{i_1} = \alpha^{i_2}$, it reveals the circular behavior of the exponents.

The main advantage of the BCH codes compared to other cyclic codes is their guaranteed lower bound on the minimum distance [21]: if $\alpha^{i_1}, \ldots, \alpha^{i_d}$ are different roots of $g(x)$ (not necessarily all the roots) such that $i_1, \ldots, i_d$ form an arithmetic progression, then $\tilde{d}_{min} \geq d + 1$.

Now we get back to our code design approach. We construct the desired code generating polynomials by investigating their parity check polynomial which is defined as:

$$h(x) \triangleq \frac{x^{2^{\tilde{m}}-1} + 1}{g(x)} \quad (11)$$

[2]BCH codes are considered due to the existence of a deterministic lower bound on their minimum distance. Although there exists a similar bound for Reed-Muller codes, they can only produce sampling matrices with small RIP order ($k$), due to their small minimum distance compared to the block size.

In other words, each field element is the root of exactly one of the $g(x)$ and $h(x)$. We construct $h(x)$ by introducing its roots. Let $l < \tilde{m} - 1$ be an integer and define

$$\mathcal{G}_{\tilde{m}}^{(l)} = \{\alpha^0, \alpha^1, \ldots, \alpha^{2^{\tilde{m}-1}+2^l-1}\} \quad (12)$$

Note that the definition of $\mathcal{G}_{\tilde{m}}^{(l)}$ depends on the choice of the primitive element ($\alpha$). We further define $\mathcal{H}_{\tilde{m}}^{(l)}$ as the subset of $\mathcal{G}_{\tilde{m}}^{(l)}$ which is closed with respect to the conjugate operation:

$$\mathcal{H}_{\tilde{m}}^{(l)} \triangleq \{r \in \mathcal{G}_{\tilde{m}}^{(l)} \mid \forall j \in \mathbb{N}: r^{2^j} \in \mathcal{G}_{\tilde{m}}^{(l)}\} \quad (13)$$

The above definition shows that if $r \in \mathcal{H}_{\tilde{m}}^{(l)}$ then its conjugate $r^{2^j}$ is also in $\mathcal{H}_{\tilde{m}}^{(l)}$. Now let us define $h(x)$:

$$h(x) = \prod_{r \in \mathcal{H}_{\tilde{m}}^{(l)}} (x - r) \quad (14)$$

As discussed before, if $r$ is a root of $h(x)$, all its conjugates are also roots of $h(x)$; therefore, $h(x) \in GF(2)[x]$, which is a required condition. Also,

$$\begin{aligned} 1 = \alpha^0 \in \mathcal{G}_{\tilde{m}}^{(l)} &\Rightarrow 1 \in \mathcal{H}_{\tilde{m}}^{(l)} \\ &\Rightarrow (1+x)|h(x) \end{aligned} \quad (15)$$

which means that the all one vector is a valid code word:

$$c = [\underbrace{1, \ldots, 1}_{2^{\tilde{m}}-1}]^T$$

$$\Rightarrow c(x) = 1 + x + \cdots + x^{2^{\tilde{m}}-2} = \frac{x^{2^{\tilde{m}}-1} + 1}{x + 1}$$

$$\Rightarrow x^{2^{\tilde{m}}-1} + 1 \big| (x^{2^{\tilde{m}}-1} + 1)\frac{h(x)}{1+x} = c(x)h(x) \quad (16)$$

Hence, the code generated by $g(x) = \frac{x^{\tilde{n}}+1}{h(x)}$ is a symmetric code and fulfills the requirement of Theorem 1. For the minimum distance of the code, note that the roots of $h(x)$ form a subset of $\mathcal{G}_{\tilde{m}}^{(l)}$; thus, all the elements in $GF(2^{\tilde{m}}) \setminus \mathcal{G}_{\tilde{m}}^{(l)}$ are the roots of $g(x)$:

$$\forall\, 2^{\tilde{m}-1} + 2^l \leq j \leq 2^{\tilde{m}} - 2: \quad g(\alpha^j) = 0 \quad (17)$$

Consequently, there exists an arithmetic progression of length $2^{\tilde{m}-1} - 2^l - 1$ among the powers of $\alpha$ in the roots of $g(x)$. As a result:

$$\tilde{d}_{min} \geq (2^{\tilde{m}-1} - 2^l - 1) + 1 = 2^{\tilde{m}-1} - 2^l \quad (18)$$

In coding theory, it is usual to look for a code with maximum $\tilde{d}_{min}$ given $\tilde{n}, \tilde{k}$. Here, we have designed a code with good $\tilde{d}_{min}$ for a given $\tilde{n}$ but with unknown $\tilde{k}$:

$$\begin{aligned} \tilde{n} &= \tilde{k} + deg(g(x)) \\ \Rightarrow \tilde{k} &= \tilde{n} - deg(g(x)) \\ &= \big(deg(g(x)) + deg(h(x))\big) - deg(g(x)) \\ &= deg(h(x)) = |\mathcal{H}_{\tilde{m}}^{(l)}| \end{aligned} \quad (19)$$

The following theorem reveals how $|\mathcal{H}_{\tilde{m}}^{(l)}|$ should be calculated.

**Theorem 2:** With the previous terminology, $|\mathcal{H}_{\tilde{m}}^{(l)}|$ is equal to the number of binary sequences of length $\tilde{m}$ such that if the sequence is written around a circle, between each two 1's, there exists at least $\tilde{m} - l - 1$ zeros.

**Proof.** We show that there exists a 1-1 mapping between the elements of $\mathcal{H}_{\tilde{m}}^{(l)}$ and the binary sequences. Let $(b_{\tilde{m}-1}, \ldots, b_0) \in \{0,1\}^{\tilde{m}}$ be one of the binary sequences and let $\beta$ be the decimal number with the binary representation that coincides with the sequence:

$$\beta = (\overline{b_{\tilde{m}-1} \ldots b_0})_2 = \sum_{i=0}^{\tilde{m}-1} b_i 2^i \quad (20)$$

We will show that $\alpha^\beta \in \mathcal{H}_{\tilde{m}}^{(l)}$. For the sake of simplicity, let us define $\beta_j$ as the decimal number that its binary representation is the same as the sequence ($\beta$) subjected to $j$ units of left circular shift ($\beta_0 = \beta$):

$$\begin{aligned}
\beta_0 &= (\overline{b_{\tilde{m}-1} \ldots b_0})_2 \\
\beta_1 &= (\overline{b_{\tilde{m}-2} \ldots b_0 b_{\tilde{m}-1}})_2 \\
\beta_2 &= (\overline{b_{\tilde{m}-3} \ldots b_0 b_{\tilde{m}-1} b_{\tilde{m}-2}})_2 \\
&\vdots \\
\beta_{\tilde{m}-1} &= (\overline{b_0 b_{\tilde{m}-1} \ldots b_1})_2
\end{aligned} \quad (21)$$

Now we have:

$$\begin{aligned}
2\beta_j &= 2 \times (\overline{b_{\tilde{m}-1-j} \ldots b_0 b_{\tilde{m}-1} b_{\tilde{m}-j}})_2 \\
&= 2^{\tilde{m}} b_{\tilde{m}-1-j} + (\overline{b_{\tilde{m}-2-j} \ldots b_0 b_{\tilde{m}-1} b_{\tilde{m}-j} 0})_2 \\
&\equiv \beta_{j+1} \pmod{2^{\tilde{m}} - 1} \\
\Rightarrow \quad & \beta_j \equiv 2^j \beta \pmod{2^{\tilde{m}} - 1} \\
\Rightarrow \quad & \alpha^{\beta_j} = \alpha^{2^j \beta}
\end{aligned} \quad (22)$$

which shows that $\{\alpha^{\beta_j}\}_j$ are conjugates of $\alpha^\beta$. To show $\alpha^\beta \in \mathcal{H}_{\tilde{m}}^{(l)}$, we should prove that all the conjugates belong to $\mathcal{G}_{\tilde{m}}^{(l)}$, or equivalently, we should show $0 \leq \beta_j \leq 2^{\tilde{m}-1} + 2^l - 1$. It is clear that $0 < \beta_j$; to prove the right inequality we consider two cases:

1) MSB of $\beta_j$ is zero:
$$b_{\tilde{m}-1-j} = 0 \Rightarrow \beta_j < 2^{\tilde{m}-1} < 2^{\tilde{m}-1} + 2^l - 1 \quad (23)$$

2) MSB of $\beta_j$ is one; therefore, according to the property of the binary sequences, the following $\tilde{m} - l - 1$ bits are zero:
$$\begin{aligned}
b_{\tilde{m}-1-j} = 1 &\Rightarrow b_{\tilde{m}-2-j} = \cdots = b_{l-j} = 0 \\
&\Rightarrow \beta_j \leq 2^{\tilde{m}-1} + \sum_{j=0}^{l-1} 2^j \\
&\Rightarrow \beta_j \leq 2^{\tilde{m}-1} + 2^l - 1 \quad (24)
\end{aligned}$$

Up to now, we have proved that each binary sequence with the above zero-spacing property can be assigned to a separate root of $h(x)$. To complete the proof, we show that if the binary representation of $\beta$ does not satisfy the property, then we have $\alpha^\beta \notin \mathcal{H}_{\tilde{m}}^{(l)}$. In fact, by circular shifts introduced in $\beta_j$, all the bits can be placed in the MSB position; thus, if the binary representation of $\beta$ does not obey the property, at least one of the $\beta_j$'s should be greater than $2^{\tilde{m}-1} + 2^l - 1$. This implies that at least one of the conjugates of $\alpha^\beta$ does not belong to $\mathcal{G}_{\tilde{m}}^{(l)}$ ∎

Theorem 2 relates the code parameter $\tilde{k}$ to a combinatorics problem. Using this relation, it is shown in Appendix A that $|\mathcal{H}_{\tilde{m}}^{(l)}| \gtrapprox \mathcal{O}\left(2^{(l+1)\frac{\ln \tilde{m}-l-1}{\tilde{m}-l-1}}\right)$.

### B. Matrix Construction

Recalling the arguments in the previous section, the choice of the polynomial $g(x)$ depends on the choice of the primitive root. In addition to this degree of freedom, from Theorem 1, no matter which code vectors from complement sets are selected, the generated matrix satisfies RIP. Hence, for a given primitive element, there are $2^{2^{\tilde{k}-1}}$ (there are $2^{\tilde{k}-1}$ complement pairs) possible matrix constructions. Among these huge number of possibilities, some have better characteristics for signal recovery from the samples. More specifically, we look for the matrices such that the columns are closed with respect to the circular shift operation: if $\mathbf{a} = [a_1, \ldots, a_{\tilde{n}}]^T$ is a column of $\mathbf{A}$, for all $1 < j \leq \tilde{n}$, $\mathbf{a}_j = [a_j, a_{j+1}, \ldots, a_{\tilde{n}}, a_1, \ldots, a_{j-1}]^T$ is also a column of $\mathbf{A}$.

The key point is that the BCH codes are a subset of cyclic codes, i.e., if $\mathbf{c}_{\tilde{n} \times 1}$ is a code vector, all its circular shifts are also valid code vectors. Thus, if we are careful in selecting from the complement sets, the generated sampling matrix will also have the cyclic property. For this selection, it should be noted that if $\mathbf{a}_{\tilde{n} \times 1}, \mathbf{b}_{\tilde{n} \times 1}$ is a complement pair and $\mathbf{c}_{\tilde{n} \times 1}$ is a circular shifted version of $\mathbf{a}_{\tilde{n} \times 1}$, the overal parity (sum of the elements in mod 2) of $\mathbf{a}_{\tilde{n} \times 1}$ and $\mathbf{b}_{\tilde{n} \times 1}$ are different (each code vector has $2^{\tilde{m}} - 1$ elements which is an odd number) while $\mathbf{a}_{\tilde{n} \times 1}$ and $\mathbf{c}_{\tilde{n} \times 1}$ have the same parity. Therefore, if we discard the code vectors with even (odd) parity (from the set of all code vectors), we are left with a set half the size of the main set such that from each complement set exactly one is selected while the set is still closed with respect to the circular shift operation. The selection algorithm is as follows:

1) For a given $k$ (compressed sensing parameter), let $i = \lceil \log_2(k) \rceil$ and choose $\tilde{m} \geq i$ (the number of compressed samples will be $m = 2^{\tilde{m}} - 1$).
2) Let $\mathcal{H}_{seq}$ be the set of all binary sequences of length $\tilde{m}$ such that 1's are circularly spaced with at least $i$ zeros. In addition, let $\mathcal{H}_{dec}$ be the set of decimal numbers such that their binary representation is a sequence in $\mathcal{H}_{seq}$.
3) Choose $\alpha$ as one of the primitive roots of $GF(2^{\tilde{m}})$ and define:
$$\mathcal{H} = \{\alpha^r \mid r \in \mathcal{H}_{dec}\} \quad (25)$$
4) Define the parity check and code generating polynomials as:
$$h(x) = \prod_{r \in \mathcal{H}} (x - r) \quad (26)$$
and
$$g(x) = \frac{x^{2^{\tilde{m}}} - 1}{h(x)} \quad (27)$$
5) Let $\tilde{\mathbf{A}}_{(2^{\tilde{m}}-1) \times (2^{deg(h)}-1)}$ be the binary matrix composed of even parity code vectors as its columns, i.e., if the columns are considered as polynomial coefficients (in $GF(2)[x]$), each polynomial should be divisible by $(x+$

| $\tilde{m}$ | $h(x)$ |
|---|---|
| 4 | $x^5 + x^4 + x^2 + 1$ |
| 6 | $x^7 + x^6 + x^2 + 1$ |
| 8 | $x^{13} + x^{12} + x^{10} + x^9 + x^8 + x^4 + x^3 + 1$ |
| 10 | $x^{26} + x^{25} + x^{24} + x^{20} + x^{16} + x^{14} + x^{13} + x^{12}$ $+ x^{10} + x^9 + x^7 + x^5 + x^4 + x^3 + x + 1$ |

TABLE I
PARITY CHECK POLYNOMIALS FOR DIFFERENT VALUES OF $\tilde{m}$ WHEN $i = 3$.

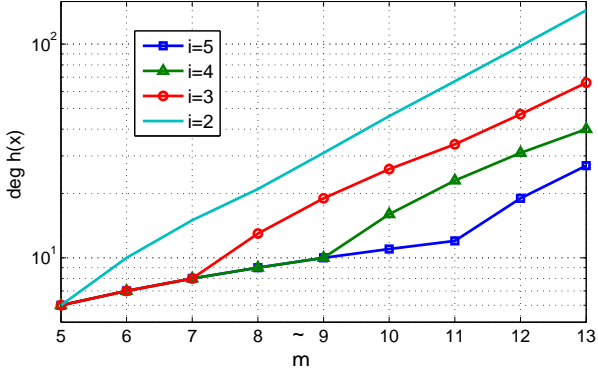

Fig. 1. Degree of $h(x)$ for different values of $\tilde{m}$ and $i$.

1) $g(x)$ (the additional factor of $x+1$ implies the even parity).
6) Replace all the zeros in $\tilde{\mathbf{A}}$ by $-1$ and normalize each column to obtain the final compressed sensing matrix ($\mathbf{A}_{(2^{\tilde{m}}-1) \times (2^{deg(h)-1})}$).

For a simple example, we consider the case $\tilde{m} = i$. It is easy to check that the number of 1's in each of the binary sequences in step 2 cannot exceed one. Therefore, we have $\mathcal{H}_{dec} = \{0, 2^0, 2^1, 2^2, \ldots, 2^{2^{i-1}}\}$. This means that $h(x)$, except for the factor $(x+1)$ is the same as the minimal polynomial of $\alpha$ (the primitive root). Since for code generation we use $(x+1)g(x)$ instead of $g(x)$, the effective $h(x)$ will be the minimal polynomial of $\alpha$ which is a primitive polynomial. In this case, the matrix $\tilde{\mathbf{A}}$ is the $(2^i-1) \times (2^i-1)$ square matrix whose columns are circularly shifted versions of the Pseudo Noise Sequence (PNS) output generated by the primitive polynomial (the absolute value of the inner product of each two columns of $\mathbf{A}$ is exactly $\frac{1}{2^i-1}$).

Table I summarizes some of the parity check polynomials for $i = 3$ (useful for $k < 8$). Also, Fig. 1 shows the degree of $h(x)$ for some of the choices of $\tilde{m}$ and $i$; the increasing rate of the degree is linear at the beginning but becomes exponential after a point.

## IV. MATRICES WITH $\{0, 1, -1\}$ ELEMENTS

We have presented a method to generate RIP-fulfilling matrices with $\pm 1$ elements. In this section, we show how binary and bipolar matrices can be combined to produce ternary matrices with larger sizes.

In order to explain the concept, we consider the $p^2 \times p^{r+1}$ binary matrices ($p$ is a prime power) in [15] where each column consists of $p$ ones (prior to normalization).

It is evident that by changing some of the 1's in the aforementioned matrix into $-1$, the norm of the columns does not change; however, the inner products change. To show how we can benefit from this feature, let us assume that $p = 2^i$; thus, there are $2^i$ nonzero elements in each column. We construct a new matrix from the original binary matrix as follows: we repeat each column $2^i$ times and then change the sign of the nonzero elements in the replicas in such a way that these nonzero elements form a Walsh-Hadamard matrix. In other words, for each column, there are $2^i$ columns (including itself) that have the same pattern of nonzero elements. The nonzero elements of these semi-replica vectors are distinct columns of the Walch-Hadamard matrix. Thus, the semi-replica vectors are orthogonal and the absolute value of the inner product of two vectors with different nonzero patterns is upper-bounded by $r$ (maximum possible value in the original matrix). Hence, the new matrix still satisfies the RIP condition with the same $k$ and $\delta_k$.

Although we have expanded the matrix with this trick, the change is negligible when the order of matrix sizes is considered ($p^2 \times p^{r+1}$ is expanded to $p^2 \times p^{r+2}$). In fact, the orthogonality of the semi-replicas is not a necessary condition; we only require that the absolute value of the inner products do not exceed $r$. This fact implies that instead of the Walsh-Hadamard matrix, we can use other $\pm 1$ matrices with more number of columns (with the same number of rows) such that their columns are almost orthogonal (inner product less than $r$). This is the case for the matrices introduced in the previous sections.

In order to mathematically describe the procedure, we need to define an operation. Let $\mathbf{s}$ be a $\beta \times 1$ binary vector with exactly $\alpha$ elements of 1 in locations $r_1, \ldots, r_\alpha \in \{1, 2, \ldots, \beta\}$. Also, let $\mathbf{x}_{\alpha \times 1} = [x_1, \ldots, x_\alpha]^T$ be an arbitrary vector. We define $\mathbf{y}_{\beta \times 1} = \mathcal{M}(\mathbf{s}, \mathbf{x})$ as:

$$\begin{cases} \forall\, 1 \leq j \leq \alpha : & y_{r_j} = x_j \\ \forall\, j \notin \{r_1, \ldots, r_\alpha\} : & y_j = 0 \end{cases} \quad (28)$$

From the above definition, we can see:

$$\langle \mathcal{M}(\mathbf{s}, \mathbf{x}_1),\, \mathcal{M}(\mathbf{s}, \mathbf{x}_2) \rangle = \langle \mathbf{x}_1, \mathbf{x}_2 \rangle \quad (29)$$

Furthermore, if the elements of both $\mathbf{x}_1, \mathbf{x}_2$ lie in the closed interval $[-1, 1]$, we have:

$$\left| \langle \mathcal{M}(\mathbf{s}_1, \mathbf{x}_1),\, \mathcal{M}(\mathbf{s}_2, \mathbf{x}_2) \rangle \right| \leq \langle \mathbf{s}_1, \mathbf{s}_2 \rangle \quad (30)$$

For the matrix construction, let $\tilde{m}$ be an integer such that $p = 2^{\tilde{m}} - 1$ is a prime (the primes of this form are called Mersenne primes). Let $k < p$ be the required order of the RIP condition and let:

$$r = \lfloor \tfrac{p}{k} \rfloor \quad,\quad i = \lceil \log_2 k \rceil \quad (31)$$

Also let $\mathbf{S}_{p^2 \times p^{r+1}} = [\mathbf{s}_1 \ \ldots \ \mathbf{s}_{p^{r+1}}]$ be the binary RIP-fulfilling matrix constructed as in [15] and $\mathbf{X}_{p \times 2^{\tilde{k}}} = [\mathbf{x}_1 \ \ldots \ \mathbf{x}_{2^{\tilde{k}}}]$ ($\tilde{k} = |\mathcal{H}_{\tilde{m}}^{(\tilde{m}-i)}|$ with the previous terminology) be the $\pm 1$ matrix introduced in the previous sections (we further normalize the columns of these matrices). We construct a new $p^2 \times (p^{r+1}.2^{\tilde{k}})$ matrix with elements in $\{0, 1, -1\}$ by combining these two matrices:

$$\mathbf{A} = [\mathcal{M}(\mathbf{s}_i, \mathbf{x}_j)]_{i,j} \quad (32)$$

Employing the same approach as used before, we show that $\mathbf{A}$ satisfies the RIP condition of order $k$, i.e., we show that the inner product of two distinct columns of $\mathbf{A}$ cannot exceed $\frac{1}{k-1}$ in absolute value while each column is normal:

$$\langle \mathcal{M}(\mathbf{s}_i, \mathbf{x}_j), \mathcal{M}(\mathbf{s}_i, \mathbf{x}_j) \rangle = \langle \mathbf{x}_j, \mathbf{x}_j \rangle = 1 \quad (33)$$

To study the inner product of $\mathcal{M}(\mathbf{s}_{i_1}, \mathbf{x}_{j_1})$ and $\mathcal{M}(\mathbf{s}_{i_2}, \mathbf{x}_{j_2})$, we consider two cases:

1) $i_1 = i_2$. In this case, since $\mathbf{s}_{i_1} = \mathbf{s}_{i_2}$, we have:

$$\begin{aligned}
\left| \langle \mathcal{M}(\mathbf{s}_{i_1}, \mathbf{x}_{j_1}), \mathcal{M}(\mathbf{s}_{i_2}, \mathbf{x}_{j_2}) \rangle \right| &= \left| \langle \mathbf{x}_{j_1}, \mathbf{x}_{j_2} \rangle \right| \\
&< \frac{1}{k-1} \quad (34)
\end{aligned}$$

2) $i_1 \neq i_2$ and therefore, $\mathbf{s}_{i_1} \neq \mathbf{s}_{i_2}$; since the elements of both $\mathbf{x}_{j_1}$ and $\mathbf{x}_{j_1}$ lie in $[-1, 1]$, we have:

$$\begin{aligned}
\left| \langle \mathcal{M}(\mathbf{s}_{i_1}, \mathbf{x}_{j_1}), \mathcal{M}(\mathbf{s}_{i_2}, \mathbf{x}_{j_2}) \rangle \right| &\leq \left| \langle \mathbf{s}_{i_1}, \mathbf{s}_{i_2} \rangle \right| \\
&< \frac{1}{k-1} \quad (35)
\end{aligned}$$

Inequalities (34) and (35) hold due to the RIP-fulfilling structure of the matrices $\mathbf{X}$ and $\mathbf{S}$. Hence, the claimed property of the inner products of the columns in $\mathbf{A}$ is proved. Consequently, $\mathbf{A}$ obeys the RIP condition of order $k$.

## V. Reconstruction from the Measurements

Matching Pursuit is one of the simplest methods for the recovery of sparse signals from linear projections. Here we show that this method can exactly recover the sparse signal from the noiseless samples.

Let $\mathbf{A}_{m \times n}$ and $\mathbf{s}_{n \times 1}$ be the sampling matrix and the $k$-sparse signal vector, respectively. The sampling process is defined by:

$$\mathbf{y}_{m \times 1} = \mathbf{A}_{m \times n} \cdot \mathbf{s}_{n \times 1} \quad (36)$$

For unique reconstruction of $\mathbf{s}_{n \times 1}$ from the samples $\mathbf{y}_{m \times 1}$, it is sufficient that the sampling matrix $\mathbf{A}_{m \times n}$ satisfies RIP of order $2k$ [8]. In this section, we show that if $\mathbf{A}_{m \times n}$ is any of the matrices discussed in Sec. II and III (including the ones in [15]) and satisfies RIP of order $2k$, the matching pursuit method can be used for perfect reconstruction. In addition, if $\mathbf{A}_{m \times n}$ has the circular structure in its columns, the computational complexity can be reduced (order of magnitude).

Let $\mathcal{S} = \{i_1, \ldots, i_k\} \subset \{1, \ldots, n\}$ be the support of $\mathbf{s}_{n \times 1}$ (nonzero locations); thus, we have:

$$\mathbf{y}_{m \times 1} = \mathbf{A} \cdot \mathbf{s} = \sum_{j=1}^{k} s_{i_j} \mathbf{a}_{i_j} \quad (37)$$

where $\mathbf{a}_i$ denotes the $i^{th}$ column in $\mathbf{A}$. In variants of the matching pursuit method, iterative approaches are used to estimate the original sparse vector $\mathbf{s}$. In these algorithms, the vector of the estimated input ($\hat{\mathbf{s}}_{n \times 1}$) is usually initialized by the all-zero vector and is updated within the iterations to reach its final values. Furthermore, a residual vector is defined as $\mathbf{r}_{m \times 1} = \mathbf{A}(\mathbf{s} - \hat{\mathbf{s}}) = \mathbf{y} - \mathbf{A}\hat{\mathbf{s}}$ which is obviously initialized by $\mathbf{y}$. In each iteration, the inner product of the residual vector with all the columns of $\mathbf{A}$ are evaluated to find the index of the maximum absolute value ($i_{max}$). Then, according to a rule, a subset of the elements of $\hat{\mathbf{s}}_{n \times 1}$ for which the indices has been chosen as the maximum value in this or the previous iterations are updated; similarly, using the new vector $\hat{\mathbf{s}}_{n \times 1}$, the residual vector is also updated and the whole procedure is repeated until a stopping condition (e.g., maximum number of iterations) is reached. Here, we show that the index of the maximum inner product at each iteration belongs to $\mathcal{S}$ (irrespective of the updating rule); this means that if a proper updating rule is used (such as in OMP), after $k$ iterations the support of $\mathbf{s}$ is completely known and therefore, perfect recovery is possible. We show this by induction: assume that up to the $t^{th}$ iteration, only the elements in $\mathcal{S}$ have appeared as indices of the maximum inner products; this means that at the beginning of the $t^{th}$ iteration, the support of $\hat{\mathbf{s}}$ and consequently the support of $\delta = \mathbf{s} - \hat{\mathbf{s}}$ are subsets of $\mathcal{S}$ (only these elements might have been updated). Without loss of generality, assume $|\delta_{i_1}| \geq |\delta_{i_2}| \geq \cdots \geq |\delta_{i_k}|$ (the rest of the $\delta_i$'s are zero). We then have:

$$\begin{aligned}
|\langle \mathbf{r}, \mathbf{a}_{i_1} \rangle| &= \left| \langle \sum_{j=1}^{k} \delta_{i_j} \mathbf{a}_{i_j}, \mathbf{a}_{i_1} \rangle \right| \\
&\geq |\delta_{i_1}| \langle \mathbf{a}_{i_1}, \mathbf{a}_{i_1} \rangle - \sum_{j=2}^{k} |\delta_{i_j}| |\langle \mathbf{a}_{i_j}, \mathbf{a}_{i_1} \rangle| \quad (38)
\end{aligned}$$

Recalling the properties of the matrix $\mathbf{A}$, we know that the columns are normal and the inner product of each two distinct columns is less than (absolute value) $\frac{1}{2k-1}$, thus,

$$\begin{aligned}
|\langle \mathbf{r}, \mathbf{a}_{i_1} \rangle| &> |\delta_{i_1}| - \frac{1}{2k-1} \sum_{j=2}^{k} |\delta_{i_j}| \\
&\geq |\delta_{i_1}| - \frac{k-1}{2k-1}|\delta_{i_1}| = \frac{k}{2k-1}|\delta_{i_1}| \quad (39)
\end{aligned}$$

On the other hand, if $l \notin \mathcal{S}$, we have:

$$\begin{aligned}
|\langle \mathbf{r}, \mathbf{a}_l \rangle| &= \left| \sum_{j=1}^{k} \delta_{i_j} \langle \mathbf{a}_{i_j}, \mathbf{a}_l \rangle \right| \\
&< \frac{1}{2k-1} \sum_{j=1}^{k} |\delta_{i_j}| \leq \frac{k}{2k-1}|\delta_{i_1}| \quad (40)
\end{aligned}$$

Combining (39) and (40), we get:

$$|\langle \mathbf{r}, \mathbf{a}_l \rangle| < \frac{k}{2k-1}|\delta_{i_1}| < |\langle \mathbf{r}, \mathbf{a}_{i_1} \rangle| \quad (41)$$

Hence, the largest inner product is obtained either with $\mathbf{a}_{i_1}$ or one of the other $\mathbf{a}_{i_j}$'s. Therefore, the index of the largest inner product is a member of the support of $\mathbf{s}_{n \times 1}$ which completes the proof for the induction. Consequently, if the matching pursuit is equipped with a proper updating rule, we expect perfect reconstruction for noiseless measurements.

As explained above, in each iteration of the matching pursuit algorithm, the inner product of $\mathbf{r}_{m \times 1}$ with all the columns in

$\mathbf{A}_{m \times n}$ needs to be calculated. Each inner product requires $m$ multiplications and $m-1$ additions. Now we show how one can benefit from the circular property of the columns of $\mathbf{A}$ to reduce the computational complexity of the reconstruction. Let $\mathbf{a}$ be one of the columns in $\mathbf{A}$ and $\mathbf{a}^{(j)}$ be its $j^{th}$ circularly shifted (to the left) version. Due to the circular property of $\mathbf{A}$, $\mathbf{a}^{(j)}$'s are all columns of $\mathbf{A}$; thus, $\langle \mathbf{a}^{(j)}, \mathbf{r} \rangle$ has to be calculated for all $j$. Let $\{\mathbf{a}^{(1)}, \mathbf{a}^{(2)}, \ldots, \mathbf{a}^{(\mu)}\}$ be different elements of $\{\mathbf{a}^{(j)}\}_j$ (obviously $\mu \leq m$ and more precisely $\mu | m$). These inner products require $\mu m$ multiplications and $\mu(m-1)$ additions if calculated directly.

A fast approach for evaluation of these values is to employ Discrete Fourier Transform (DFT) or its fast implementation -FFT. The key point in this approach is that the inner products can be found through circular convolution of $\mathbf{r}$ and $\mathbf{a}$, i.e.,

$$\langle \mathbf{r}, \mathbf{a}_{(j)} \rangle = \left( \mathbf{r} \circledast_m \mathbf{a} \right)\big|_j \quad (42)$$

where $\circledast_m$ represents the circular convolution with period $m$. It is well-known that the circular convolution can be easily calculated using DFT: if $\mathbf{r}_f$ and $\mathbf{a}_f$ denote the DFT of $\mathbf{r}$ and $\mathbf{a}$, respectively, we have:

$$IDFT\{\mathbf{r}_f \odot \mathbf{a}_f\} = \left[ \left( \mathbf{r} \circledast_m \mathbf{a} \right)\big|_0, \ldots, \left( \mathbf{r} \circledast_m \mathbf{a} \right)\big|_{m-1} \right] \quad (43)$$

where $\mathbf{v}_{m \times 1} \odot \mathbf{u}_{m \times 1} \triangleq [v_1 u_1, \ldots, v_m u_m]^T$. For evaluation of the inner products by this method, $\mathbf{r}_f$ has to be calculated only once (at each iteration) using DFT. Thus, excluding the calculation of $\mathbf{r}_f$ (which is done only once), the inner products of $\mathbf{r}$ with $\{\mathbf{a}^{(j)}\}_j$ require one $DFT$, one $IDFT$ and $m$ multiplications. Since $\mu$ different circular shifts of $\mathbf{a}$ are possible, at most $\mu$ coefficients of $\mathbf{a}_f$ at equi-distant positions are nonzero; hence, $\mu$-point DFT (and consequently IDFT) of $\mathbf{a}_{m \times 1}$ rather than the general $m$-point DFT is adequate. For $\mu$-point DFT of $\mathbf{r}$, we can simply down-sample the evaluated $m \times 1$ vector of $\mathbf{r}_f$ (note that $\mu | m$) and there is no need for an extra $\mu$-point DFT. Employing the FFT version, we require $2\mu \lceil \log_2 \mu \rceil$ multiplications and $m - \mu + \mu \lceil \log_2 \mu \rceil$ additions per $\mu$-point DFT or IDFT. Comparison of the number of required multiplications in calculation of the above $\mu$ inner products reveals the efficiency of the DFT approach; i.e., less computational complexity is required for reconstruction of the signal from the measurements obtained from a sensing matrix with circular property in the columns. It should be emphasized that by using the FFT method, the reconstruction algorithm and therefore, the results are essentially the same (matching pursuit), however, due to the circular format of the columns in the sensing matrix, the required computational complexity is significantly reduced.

## VI. Numerical Results

For simulation results we have investigated the binary ,bipolar and ternary matrices which satisfy RIP of order $k = 4$. For binary matrices, the Devore's structure in [15] with the size of $64 \times 512$ is considered while for bipolar matrices using BCH codes, the matrix size is $63 \times 512$. For the ternary matrix, we used the mixture of the Devore's binary $49 \times 343$ and the bipolar $7 \times 8$ matrices that both satisfy the RIP order of at least $k = 4$; the final matrix in this fashion would be a ternary $49 \times 2744$ matrix which satisfies the RIP of at least $k = 4$, however, we keep only the first $512$ columns of this matrix to have the same $n$ in all cases. In addition, we have included two Gaussian random matrices as representatives of the random compressed sensing, one with the size $64 \times 512$ and the other with $49 \times 512$. After generating the matrices, we observed the coherence (maximum inner product between distinct columns) values $\frac{1}{7}$, $\frac{1}{4}$ and $\frac{1}{4}$ for the bipolar, binary and ternary matrices, respectively; i.e., although the bipolar matrix is designed for the RIP of order 4, the orders up to and including 7 are also guaranteed.

Figure 2 shows the percentage (probability) of perfect recovery (SNR$_{rec.} \geq 100dB$) when different sparsity orders are considered. For the generation of the $k$-sparse input signals in the simulations, we first select the support (nonzero locations) uniformly at random among the $\binom{512}{k}$ possibilities and then generate the corresponding values by realizations of $k$ normal random variables; furthermore, the depicted percentage curves are found by averaging the results for 5000 different input signals (for each $k$). For the reconstruction of the $k$-sparse input signals from the compressed measurements (noisy or noiseless), we perform $k$ steps (we assume $k$ is known for the decoder) of the Orthogonal Matching Pursuit (OMP); i.e., irrespective of the residual vector, we perform $k$ OMP iterations. We chose OMP to benefit from the small coherence of the matrices; OMP is much faster than the basis pursuit method (and its variants like SPGL1) in our setup. As shown in Fig. 2, for $k = 4$ all of the deterministic matrices are able to recover the sparse signal, and their performance degrade when $k$ increases (the matrices are designed for $k = 4$). An interesting observation is that the bipolar matrix outperforms other matrices when larger $k$'s are considered; for example, at $k = 20$, the perfect recovery percentage using the samples obtained from the bipolar matrix is almost $24\%$ better on average than that of the random matrix and $29\%$ better than that of the Devore's Matrix. Although it seems that the ternary matrix falls short of the performance, it should be reminded that it uses fewer number of samples (rows) and also it supports far more number of columns (we chose 512 of the total 2744). As can be observed, this matrix outperforms the random matrix of the similar size.

In order to include the noise effect in our results, in Fig. 3, we have considered the recovery using the same matrices at $k = 15$ (an overloaded value) when various noise levels are accompanied with the measurements ($\mathbf{y}$). Again the same OMP method is employed for the reconstruction and the results are averaged over 10000 runs. The results confirm that the performance curve is continuous (stability) when noise is included. Since the curves for all the matrices coincide for $k = 4$, we did not include it.

## VII. Conclusion

In this paper, we introduced a new connection between the OOC codes and RIP fulfilling matrices which results in the construction of binary sampling matrices. We have further presented a design for bipolar matrices using linear binary correction codes, especially BHC codes. In the latter design, we replace the zeros in the binary linear code vectors by $-1$ and use them as the columns of the sensing matrix. These matrices, in addition to their deterministic and known structure, have two main advantages: 1) simplicity of the measurement process; real/complex entries in the sensing matrix increases the computational complexity of the sampler as well as the required bit-precision for storing the samples, and 2) simplicity of the reconstruction process. Due to the cyclic property of the columns inherited from the cyclic codes, the FFT algorithm can speed up the reconstruction procedure. These $\pm 1$ matrices are further expanded by considering $\{0, 1, -1\}$ elements; this expansion is achieved by combining the bipolar and binary matrices. Although the generated matrices show an improvement in the realizable size of the RIP-constrained matrices, the bound predicted by random matrices cannot be achieved.

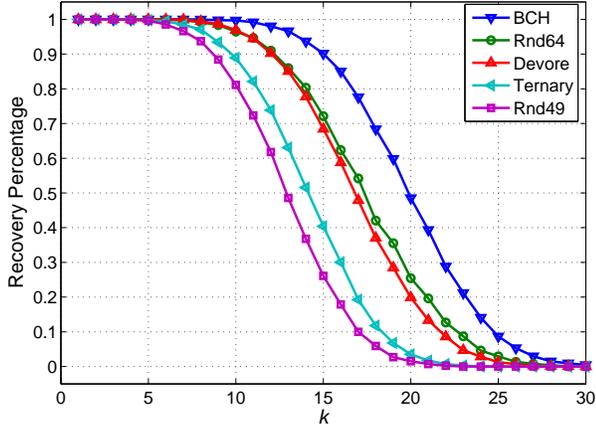

Fig. 2. The recovery percentage ($SNR_{rec.} \geq 100dB$) for different sparsity values ($k$). Sampling matrices for BCH, Devore and ternary methods satisfy RIP of order at least 4.

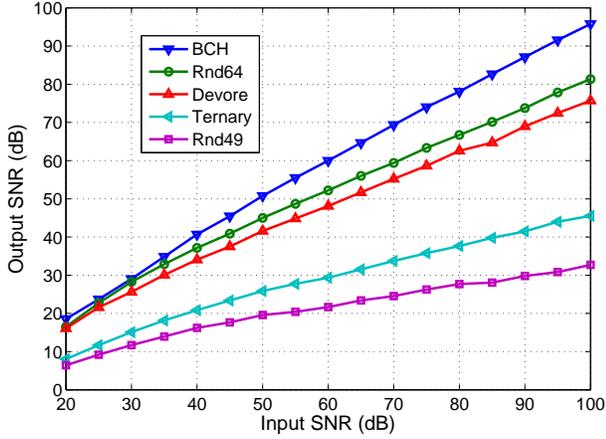

Fig. 3. The SNR of the reconstructed signal for 15-sparse signals when the compressed samples are accompanied with different noise powers. Sampling matrices for BCH, Devore and ternary methods satisfy RIP order at least 4.

## APPENDIX A

In Theorem 2, we showed that $\tilde{k}$ is equal to the number of binary sequences of length $\tilde{m}$ such that no two 1s are spaced by less than $\tilde{m} - l - 1$ zeros (circular definition). To evaluate this number, let us define $\tau_b^{(a)}$ as the number of binary sequences of length $b$ such that if the sequence is put around a circle, between each two 1's, there is at least $a$ zeros. In addition, let $\kappa_b^{(a)}$ be the number of binary sequences such 1's are spaced by at least $a$ zeros apart (circular property is no longer valid for $\kappa_b^{(a)}$). We first calculate $\kappa_b^{(a)}$ and then we show the connection between $\kappa_b^{(a)}$ and $\tau_b^{(a)}$.

There are two kinds of binary sequences counted in $\kappa_b^{(a)}$:
1) The last bit in the sequence is 0; by omitting this bit, we obtain a sequence of length $b-1$ with the same property. Also, each binary sequence of length $b-1$ with the above property can be padded by 0 while still satisfying the required property to be included in $\kappa_b^{(a)}$. Therefore, there are $\kappa_{b-1}^{(a)}$ binary sequence of this type.
2) The last bit in the sequence is 1; this means that the last $a+1$ bits of the sequence are $\underbrace{0, \ldots, 0}_{a}, 1$. Similar to the above case, each binary sequence of length $b - a - 1$ counted in $\kappa_{b-a-1}^{(a)}$ can be padded by the block $\underbrace{0, \ldots, 0}_{a}, 1$ to produce a sequence included in $\kappa_b^{(a)}$. Thus, there are $\kappa_{b-a-1}^{(a)}$ binary sequences of this type.

In summary, we have the following recursive equation:

$$\kappa_b^{(a)} = \kappa_{b-1}^{(a)} + \kappa_{b-a-1}^{(a)} \tag{44}$$

Since for $b \leq a + 1$, there can be at most one 1 in the binary sequence, we thus have:

$$1 \leq b \leq a+1: \kappa_b^{(a)} = b + 1 \tag{45}$$

From (44), the last initial condition ($\kappa_{a+1}^{(a)} = a + 2$) is equivalent to $\kappa_0^{(a)} = 1$. If we define the onesided $\mathcal{Z}$-transform of $\kappa_b^{(a)}$ as follows

$$\kappa^{(a)}(z) = \sum_{b=0}^{\infty} \kappa_b^{(a)} z^{-b}, \tag{46}$$

it is not hard to check that:

$$\kappa^{(a)}(z) = \frac{1}{1 - z^{-1}} \cdot \frac{1 - z^{-(a+1)}}{1 - z^{-1} - z^{-(a+1)}} \tag{47}$$

Therefore, the increasing rate $\kappa_b^{(a)}$ with respect to $b$ ($b \gg 1$) has the same order as $\gamma^b$ where $\gamma$ is the largest (in absolute value) root of $f(z) = z^{a+1} - z^a - 1$. Since $f(1) \cdot f(2) < 0$, there is a real root in $(1, 2)$; let us denote this root by $\gamma$. In fact, $\gamma$ is the largest root of $f(z)$ (we do not prove this; however, if $f(z)$ has a larger root, the increasing rate of $\kappa_b^{(a)}$ would be greater than $\gamma^b$):

$$1 < \gamma < 2, \quad f(\gamma) = \gamma^{a+1} - \gamma^a - 1 = 0 \tag{48}$$

Since $\gamma > 1$ we can assume $\gamma = 1 + \frac{1}{\delta}$, where $\delta > 1$:

$$\gamma^{a+1} - \gamma^a = 1 \Rightarrow \left(1 + \frac{1}{\delta}\right)^a = \delta \tag{49}$$

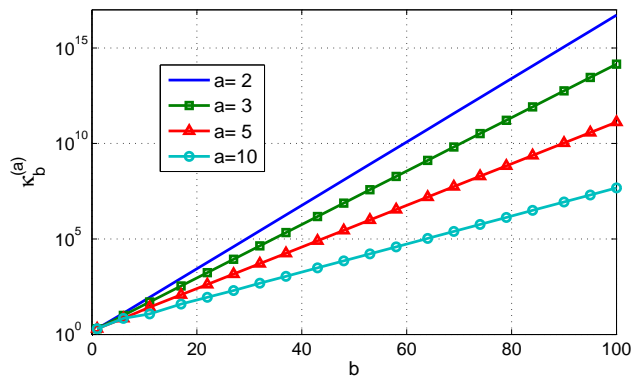

Fig. 4. Exact values of $\kappa_b^{(a)}$ for different values of $a$ and $b$.

In appendix B we show that $\delta > a^{0.7}$, thus, we have:

$$\gamma^{\frac{a}{\ln a}} = \left(1 + \frac{1}{\delta}\right)^{\frac{a}{\ln a}} = \delta^{\frac{1}{\ln a}} > a^{\frac{0.7}{\ln a}} = e^{0.7} > 2 \quad (50)$$

Now we can show the connection between $\tau_b^{(a)}$ and $\kappa_b^{(a)}$. According to the definition of these parameters, we see that every binary sequence counted in $\tau_b^{(a)}$ is also counted in $\kappa_b^{(a)}$, therefore, $\tau_b^{(a)} \leq \kappa_b^{(a)}$. In addition, if a sequence counted in $\kappa_{b-a}^{(a)}$ is padded with $a$ zeros at the end, it satisfies the requirements to be counted in $\tau_b^{(a)}$, thus, $\kappa_{b-a}^{(a)} \leq \tau_b^{(a)}$. Combining the latter two inequalities, we get:

$$\mathcal{O}(\gamma^{b-a}) \leq \tau_b^{(a)} \leq \mathcal{O}(\gamma^b) \quad (51)$$

The above equation in conjunction with the result in (50), yields:

$$\left(\tau_b^{(a)}\right)^{\frac{a}{\ln a}} \gtrapprox \mathcal{O}(2^{b-a}) \quad (52)$$

The interpretation of the above inequality for $\tilde{k}$ is as follows:

$$\tilde{k} = \tau_{\tilde{m}}^{(\tilde{m}-l-1)} \gtrapprox \mathcal{O}\left(2^{(l+1)\frac{\ln \tilde{m}-l-1}{\tilde{m}-l-1}}\right) \quad (53)$$

Note that $\tilde{m} - l - 1 \leq \log_2 k$ where $k$ is the maximum RIP order that we can guarantee by the arguments in this paper. Hence, for the constructed bipolar matrices we have:

$$m = 2^{\tilde{m}} - 1 < 2^{\tilde{m}-l-1} 2^{l+1} \lessapprox \mathcal{O}\left(k\bigl(\log_2 n\bigr)^{\frac{\log_2 k}{\ln \log_2 k}}\right) \quad (54)$$

Figure 4 shows the asymptotic behavior of $\kappa_b^{(a)}$ at different $a$ values when $b$ increases. For comparison, Fig. 4 suggests that $\kappa_b^{(5)} \approx 1.66 \times 1.285^b$ while our approximation is $\kappa_b^{(5)} \geq \mathcal{O}(1.25^b)$.

## APPENDIX B

In this appendix we show that if $\left(1 + \frac{1}{\delta}\right)^a = \delta$, where $a \in \mathbb{N}$ and $1 < \delta \in \mathbb{R}$, then $\delta > a^{0.7}$. To prove this claim, we start by the following lemma:

*Lemma 1:* The function $f(x) = x^{0.3} - 0.5 x^{-0.4} - 0.7 \ln x$ takes only positive values for $x > 0$.

*Proof* It is obvious that $\lim_{x \to +\infty} f(x) = +\infty$. We prove the lemma by showing that $f'(x)$ (derivative of $f$) has only one root in $[0, \infty)$ which gives the minimum value of $f$:

$$\begin{aligned} f'(x) &= 0.3 x^{-0.7} + 0.2 x^{-1.4} - 0.7 x^{-1} \\ &= \frac{3(x^{0.1})^7 - 7(x^{0.1})^4 + 2}{10 x^{1.4}} \end{aligned} \quad (55)$$

The only positive root of the polynomial $3y^7 - 7y^4 + 2$ is $y \approx 1.277$; thus, the minimum value of $f(x)$ on the positive axis is attained at $x \approx 1.277^{10} \approx 11.532$. Evaluation of the function at this point shows $f(11.532) \approx 0.18 > 0$ ■

Now by using the above lemma, for $x \geq 1$ we have:

$$\begin{aligned} 0 &< f(x) = x\bigl(x^{-0.7} - 0.5 x^{-1.4}\bigr) - \ln x^{0.7} \\ &\leq x \ln\bigl(1 + x^{-0.7}\bigr) - \ln x^{0.7} \end{aligned} \quad (56)$$

Therefore,

$$e^{\ln x^{0.7}} < e^{x \ln\left(1 + x^{-0.7}\right)} \quad \Rightarrow \quad \frac{\left(x^{0.7}\right)^{x+1}}{\left(1 + x^{0.7}\right)^x} < 1 \quad (57)$$

Note that the function $\psi(x) = \frac{x^{a+1}}{(1+x)^a} = x\left(1 - \frac{1}{1+x}\right)^a$ is strictly increasing on the positive axis (both $x$ and $1 - \frac{1}{1+x}$ are increasing). Recalling the equality $\left(1 + \frac{1}{\delta}\right)^a = \delta$, we know $\psi(\delta) = 1$; on the other hand, if we put $x = a$ in (57) we get $\psi\bigl(a^{0.7}\bigr) < 1$ which reveals that $\delta > a^{0.7}$.

## ACKNOWLEDGMENT

The authors sincerely thank K. Alishahi for his help in the proof given in the appendix and A. Hoseini, M. Sharif and P. Pakrouh for their support with the simulation results. We are indebted to M. Alem from ONR lab. at Sharif University of Technology for his assistance in the OOC part and V. Montazerhodjat for his fruitful comments. We also acknowledge the constructive comments of J. Romberg which highly improved the introduction section.